\documentclass[12pt]{article}

\usepackage{amsthm,amsmath,amssymb}
\usepackage{graphicx}

\usepackage[linesnumbered,ruled,vlined]{algorithm2e}
\usepackage{dsfont}

\newtheorem{theorem}{Theorem}
\newtheorem{lemma}[theorem]{Lemma}

\newcommand{\OO}{{\mathcal{O}}}
\newcommand{\m}[1]{\mathrm{#1}}
\newcommand{\classNP}{\mathcal{NP}}
\newcommand{\classP}{\mathcal{P}}

\title{\bf Improved Approximation Algorithm for the Number of Queries Necessary to Identify a Permutation}
\author{Mourad El Ouali \qquad Volkmar Sauerland\\
\small Department of Computer Science\\[-0.8ex]
\small Christian-Albrechts-Universit\"{a}t zu Kiel\\[-0.8ex]
\small Kiel, Germany\\
\small\tt <meo,vsa>@informatik.uni-kiel.de\\
}
\date{\today}

\begin{document}

\maketitle

\begin{abstract}
In the past three decades, deductive games have become interesting from the algorithmic point of view.
Deductive games are two players zero sum games of imperfect information.
The first player, called ${"}$codemaker${"}$, chooses a secret code and the second player, called ${"}$codebreaker${"}$, tries to break the secret code by making as few guesses as possible, exploiting information that is given by the codemaker after each guess.
A well known deductive game is the famous Mastermind game.
In this paper, we consider the so called Black-Peg variant of Mastermind, where the only information concerning a guess is the number of positions in which the guess coincides with the secret code.
More precisely, we deal with a special version of the Black-Peg game with $n$ holes and $k\ge n$ colors where no repetition of colors is allowed.
We present a strategy that identifies the secret code in $\OO(n\log_2{n})$ queries.
Our algorithm improves the previous result of Ker-I Ko and Shia-Chung Teng (1985) by almost a factor of 2 for the case $k=n$.
To our knowledge there is no previous work dealing with the case $k>n$.

\end{abstract}

\section{Introduction}

In the past three decades, deductive games have become interesting from the algorithmic point of view.
In this kind of games, two players are involved.
They are called the codemaker and the codebreaker, respectively.
One of the most famous games of this kind is Mastermind.

\subsection{Problem Description}
Mastermind is a two players board game invented in 1970 by the postmaster and telecommunication expert Mordecai Meirowitz.
The original version of Mastermind consists of a board with twelve (or ten, or eight) rows containing four holes and pegs of six different colors.
The idea of the game is that the codemaker, say ${"}$Carole${"}$, chooses a secret color combination of four pegs from the six possible colors and the codebreaker, say ${"}$Paul${"}$, has to identify the code by a sequence of queries and corresponding information that is provided by Carole.
All queries are also color combinations of four pegs.
Information is given about the number of correctly positioned colors and further correct colors, respectively.
Mathematically, we have two arbitrary positive integers, $n$ and $k$.
Carole selects a vector $y\in \{1,\dots,k\}^n$ and Paul gives in each iteration a query in form of a vector $x\in \{1,\dots,k\}^n$.
Carole replies with a pair of two numbers:
\begin{itemize}
\item A black information $\m{black}(x,\,y)$, which is the number of positions in which both vectors $x$ and $y$ coincide:
\begin{equation*}
\m{black}(x, y)=|\{i\in \{1,\dots,n\};\, x(i)=y(i) \}|.
\end{equation*}
\item A white information $\m{white}(x,\,y)$, which is the number of additional pegs with a right color but a wrong position:
\begin{equation*}
\m{white}(x, y)=\max_{\sigma\in S_{n}}|\{i\in \{1,\dots,n\};\, y(i)=x(\sigma(i)) \}|- \m{black}(x, y).
\end{equation*}
\end{itemize}
The Black-Peg game is a special version of Mastermind, where Carole answers only with the $\m{black}$ information.
A further version is the so-called AB game in which all colors within a code must be distinct.
In this paper, we deal with a special combination of the Black-Peg game and the AB game, where both the secret vector and the guesses are permutations ($k\geq n$) and the answers are given by the $\m{black}$ information, only.

\subsection{Related Works}
It is not only the playful nature of Mastermind that has attracted the attention of computer scientists, but
more importantly its relation to information-theoretic and fundamental complexity questions.
In 1963, several years before the invention of Mastermind as a commercial board game, Erd\"os and R\'enyi \cite{ER63} analyzed the same problem with two colors.
One of the earliest analysis of this game after its commercialization dealing with the case of 4 pegs and 6 colors was done by Knuth \cite{K77}.
He presented a strategy that identifies the secret code in at most 5 guesses.
Ever since the work of Knuth the general case of arbitrary many pegs and colors has been intensively investigated in combinatorics and computer science literature.

In the field of complexity, Stuckman and Zhang \cite{SZ06} showed that it is ${\classNP}$-complete to determine if a sequence of queries and answers is satisfiable.
Concerning the approximation aspect, there are many works regarding different methods \cite{BGL09,CCH96, C83, DW12, FL10, GCG11,GMC11,Goo09a,JP09, KC03, KL93, SZ06, TK03}.

The Black-Peg game was first introduced by Chv\'atal for the case $k=n$.
He gave a deterministic adaptive strategy that uses $2n\lceil\log_2{k}\rceil + 4n$ guesses.
Later, Goodrich \cite{Goo09b} improved the result of Chv\'atal for arbitrary $n$ and $k$ to $n\lceil\log_2{k}\rceil + \lceil (2-1/k)n \rceil + k$ guesses.
Moreover, he proved in same paper that this kind of game is $\classNP$-complete.
A further improvement to $n\lceil\log_2{n}\rceil + k - n +1$ for $k>n$ and $n\lceil\log_2{n}\rceil + k$ for $k\le n$ was done by J\"ager and Peczarski \cite{JP11}.
Recently, Doerr et al \cite{DSTW13} improved the result obtained by Chv\'atal to $\OO(n \log \log n)$ and also showed that this asymptotic order even holds for up to $n^2\log\log n$ colors, if both black and white information is allowed.

Another variant of Mastermind is the so-called AB game, also known as ${"}$Bulls and Cows${"}$ game.
Here, all pegs in the secret code as well as in each query must have distinct colors.

Concerning the combination of both variants, Black-Peg game and AB game, there is only one work due to Ker-I Ko and Shia-Chung Teng \cite{KT86} for the case $k=n$.
They presented a strategy that identifies the secret permutation in at most $2n\log_2{n} + 7n$ guesses and proved that the corresponding counting problem is $\# {\classP}$-complete.
To our knowledge there is no result for the case $k > n$, yet.

\subsection{Our Contribution}
In this paper we consider the Black-Peg game without color repetition.
We present a polynomial-time algorithm that identifies the secret permutation in less than $n\log_2{n}+\lambda n$ queries in the case $k=n$ and in less than $n\log_2{n}+k+2n$ queries in the case $k>n$.
Our performance in the case $k=n$ is an improvement of the result of Ker-I Ko and Shia-Chung Teng \cite{KT86} by almost a factor of 2.
Note, that additional difficulty compared to the case considered by Doerr et al is given here by the fact that color repetition is not only forbidden for the secret code but also for the guesses.

\section{An Algorithm for Permutation-Mastermind}\label{section:KequalsN}

We first consider the case $k=n$ and forbidden color repetition, meaning that codes are one-line representations of permutations in $S_n$.
For convenience, we will use the term permutation for both, a mapping in $S_n$ and its one-line representation as a vector.
Our algorithm for finding the secret permutation $y\in S_n$ includes two main phases which are based on two ideas.
In the first phase we guess an initial sequence of $n$ permutations that has a predefined structure.
In the second phase, the structure of the initial sequence and the corresponding information by the codemaker enable us to identify correct components $y_i$ of the secret code one after another, each by using a binary search.
Recall, that for two codes $w=(w_1,\dots,w_n)$ and $x=(x_1,\dots,x_n)$, we denote by $\m{black}(w,x)$ the number $|\{i\in\{1,\dots,n\}\,|\,w_i=x_i\}|$ of components in which $w$ and $x$ are equal.
We denote the mapping $x$ restricted to the set $\{s,\dots,l\}$ with $(x_i)_{i=s}^{l}$, $s,l\in \{1,\dots,n\}$.

\subsection{Phase 1}
Consider the $n$ permutations, $\sigma^1,\dots,\sigma^n$, that are defined as follows: $\sigma^1$ corresponds to the identity map and for $j\in\{1,\dots,n-1\}$, we obtain $\sigma^{j+1}$ from $\sigma^{j}$ by a circular shift to the right, i.e. we set
\begin{eqnarray}\label{initialPermutations}
\sigma^1 &=&(1,2,\dots,n),\\ \notag
\sigma^2 &=&(n,1,2,\dots,n-1),\\ \notag
\sigma^3 &=&(n-1,n,1,2,\dots,n-2),\\ \notag
& \dots,\\ \notag
\sigma^{n-1} &=&(3,4,\dots,n,1,2),\\ \notag
\sigma^n &=&(2,3,\dots,n,1).
\end{eqnarray}
Within those $n$ permutations, every color appears exactly once at every position and, thus, we have
\begin{equation}\label{infosum}
\sum_{j=1}^{n}\m{black}(\sigma^j,y)=n.
\end{equation}
We guess $\sigma^1,\dots,\sigma^{n-1}$ and obtain the additional information $\m{black}(\sigma^n,y)$ from (\ref{infosum}).

\subsection{Phase 2}
The strategy of the second phase identifies the values of $y$ one after another.
This is done by using two binary search routines, called {\sc findFirst} and {\sc findNext}, respectively.
The idea behind both binary search routines is to exploit the information that for $1\leq i,j\leq n-1$ we have $\sigma^{j}_{i}=\sigma^{j+1}_{i+1}$, $\sigma^{n}_{i}=\sigma^{1}_{i+1}$, $\sigma^{j}_{n}=\sigma^{j+1}_{1}$ and $\sigma^{n}_{n}=\sigma^{1}_{1}$.
While, except for an unfrequent special case, {\sc findFirst} is used to identify the first correct component of the secret code, {\sc findNext} identifies the remaining components in the main loop of the algorithm.
Actually, {\sc findFirst} would also be able to find the remaining components but requires more guesses than {\sc findNext} (twice as much in the worst case).
On the other hand, {\sc findNext} does only work if at least one value of $y$ is already known such that we have to identify the value of one secret code component in advance.
\subsubsection{Identifying the First Component}
Equation (\ref{infosum}) implies that either $\m{black}(\sigma^j,y)=1$ holds for all $j\in\{1,\dots,n\}$ or that we can find a $j\in\{1,\dots,n\}$ with $\m{black}(\sigma^j,y)=0$.

In the first case, which is unfrequent, we can find one correct value of $y$ by guessing at most $\frac{n}{2}+1$ modified versions of some initial guess, say $\sigma^1$.
Namely, if we define a guess $\sigma$ by swapping a pair of components of $\sigma^1$, we will obtain $\m{black}(\sigma,y)=0$, if and only if one of the swapped components has the correct value in $\sigma^1$.

In the frequent second case, we find the first component by {\sc findFirst} in at most $2\lceil\log_2{n}\rceil$ guesses.
The routine {\sc findFirst} is outlined as Algorithm \ref{findFirst} and works as follows:
In the given case, we can either find a $j\in\{1,\dots,n-1\}$ with $\m{black}(\sigma^j,y)>0$ but $\m{black}(\sigma^{j+1},y)=0$ and set $r:=j+1$, or we have $\m{black}(\sigma^n,y)>0$ but $\m{black}(\sigma^1,y)=0$ and set $j:=n$ and $r:=1$.
We call such an index $j$ an {\em active} index.
Now, for every $l\in\{2,3,\dots,n\}$ we define the code
\[
\sigma^{j,l}:=\left((\sigma^{j}_i)_{i=1}^{l-1},\sigma^{r}_{1},(\sigma^{r}_{i})_{i=l+1}^n\right),
\]
and call the peg at position $l$ in $\sigma^{j,l}$ the pivot peg.
From the information $\sigma^{j}_{i}=\sigma^{r}_{i+1}$ for $1\leq i\leq n-1$ we conclude that $\sigma^{j,l}$ is actually a new permutation as required.
The fact that $\m{black}(\sigma^{r},y)=0$ implies that the number of correct pegs up to position $l-1$ in $\sigma^{j}$ is either $\m{black}(\sigma^{j,l},y)$ (if $y_l\ne \sigma^{r}_{1}$) or $\m{black}(\sigma^{j,l},y)-1$ (if $y_l= \sigma^{r}_{1}$).
For our algorithm, we will only need to know if there exist one correct peg in $\sigma^{j}$ up to position $l-1$.
The question is cleared up, if $\m{black}(\sigma^{j,l},y) \ne 1$.
On the other hand, if $\m{black}(\sigma^{j,l},y) = 1$, we can define a new guess $\rho^{j,l}$ by swapping the pivot peg with a wrong peg in $\sigma^{j,l}$.
We define
\[
\rho^{j,l}:=\begin{cases}
\left((\sigma^{j}_{i})_{i=1}^{l},\sigma^{r}_{1},(\sigma^{r}_{i})_{i=l+2}^{n}\right)& \text{if }l<n\\
\left(\sigma^{r}_{1},(\sigma^{j}_{i})_{i=2}^{n-1},\sigma^{j}_{1}\right)& \text{if }l=n
\end{cases}
\]
assuming for the case $l=n$, that we know that $\sigma^{j}_1\ne y_1$.
We will obtain $\m{black}(\rho^{j,l},y)>0$, if and only if the pivot peg had a wrong color before, meaning that there is one correct peg in $\sigma^{j}$ in the first $l-1$ places.
Thus, we can find the position $m$ of the left most correct peg in $\sigma^{j}$ by a binary search as outlined in Algorithm \ref{findFirst}.

\begin{algorithm}
\SetKwInOut{Input}{input}\SetKwInOut{Output}{output}
\Input{Code $y$ and an active index $j\in\{1,\dots,n\}$}
\Output{Position $m$ of the left most correct peg in $\sigma^j$}
\lIf{$j=n$}{$r:=1$ }\lElse{$r:=j+1$}\;
$a:=1$\;
$b:=n$\;
$m:=n$ \tcp*{correct position in $\sigma^{j}$ to be determined}
\While{$b>a$}
{
    $l:=\lceil \frac{a+b}{2} \rceil$ \tcp*{current pivot position}
	Guess $\sigma^{j,l}:=\left((\sigma^{j}_i)_{i=1}^{l-1},\sigma^r_1,(\sigma^r_i)_{i=l+1}^{n}\right)$\;
	$s:=\m{black}(\sigma^{j,l},y)$\;
	\If{$s=1$}
	{
		 \lIf{$l<n$}{$\rho^{j,l}:=\left((\sigma^{j}_{i})_{i=1}^{l},\sigma^{r}_{1},(\sigma^{r}_{i})_{i=l+2}^{n}\right)$}\;
		\lElse{$\rho^{j,l}:=\left(\sigma^{r}_{1},(\sigma^{j}_{i})_{i=2}^{n-1},\sigma^{j}_{1}\right)$}\;
		Guess $\rho^{j,l}$\;
		$s:=\m{black}(\rho^{j,l},y)$\;
	}
	\If{$s>0$}
	{
		$b:=l-1$\;
		\lIf{$b<m$}{$m:=b$}\;
	}
	\lElse{$a:=l$}\;
}
Return $m$\;
\caption{Function {\sc findFirst}}\label{findFirst}
\end{algorithm}

\subsubsection{Identifying a Further Component}
For the implementation of {\sc findNext} we deal with a partial solution vector $x$ that satisfies $x_i\in\{0,y_i\}$ for all $i\in\{1,\dots,n\}$.
We call the (indices of the) non-zero components of the partial solution {\em fixed}.
They indicate the components of the secret code that have already been identified.
The (indices of the) zero components are called {\em open}.
Whenever {\sc findNext} makes a guess $\sigma$, it requires to know the number of open components in which the guess coincides with the secret code, i.e. the number
\[
\m{black}(\sigma,y,x) := \m{black}(\sigma,y) - \m{black}(\sigma,x).
\]
Note, that the term $\m{black}(\sigma,x)$ is known by the codebreaker.
After the first component of $y$ has been found and fixed in $x$, there exists a $j\in \{1,\dots,n\}$ such that $\m{black}(\sigma^j,y,x)=0$.
As long as we have open components in $x$, we can either find a $j\in\{1,\dots,n-1\}$ with $\m{black}(\sigma^j,y,x)>0$ but $\m{black}(\sigma^{j+1},y,x)=0$ and set $r:=j+1$, or we have $\m{black}(\sigma^n,y,x)>0$ but $\m{black}(\sigma^1,y,x)=0$ and set $j:=n$ and $r:=1$.
Again, we call such an index $j$ an {\em active} index.

\begin{algorithm}
\SetKwInOut{Input}{input}\SetKwInOut{Output}{output}
\Input{Code $y$, partial solution $x\ne 0$ and an active index $j\in\{1,\dots,n\}$}
\Output{Position $m$ of a correct open component in $\sigma^j$}
\lIf{$j=n$}{$r:=1$ }\lElse{$r:=j+1$}\;
Choose a color $c$ with identified position (a value $c$ of some non-zero component of $x$)\;
Let $l_j$ and $l_r$ be the positions with color $c$ in $\sigma^j$ and $\sigma^r$, respectively\;
\lIf{$l_j=n$}{$\mathrm{leftSearch:=true}$}
\Else
{
    Guess
	$\sigma^{j,0}:=\left(c,(\sigma^{j}_{i})_{i=1}^{l_j-1},(\sigma^{j}_{i})_{i=l_j+1}^{n}  \right)$\;
	$s:=\m{black}(\sigma^{j,0},y,x)$\;
	\lIf{$s=0$}{$\mathrm{leftSearch:=true}$}\;
	\lElse{$\mathrm{leftSearch:=false}$}\;
}
\lIf{$\mathrm{leftSearch}$}{let $a:=1$ and $b:=l_j$}\;
\lElse{let $a:=l_r$ and $b:=n$}\;
$m:=n$ \tcp*{correct position in $\sigma^{j}$ to be determined}
\While{$b>a$}
{	
$l:=\lceil \frac{a+b}{2} \rceil$ \tcp*{current position for peg $c$}
\lIf{$\mathrm{leftSearch}$}{$\sigma^{j,l}:=\left((\sigma^{j}_{i})_{i=1}^{l-1},c,(\sigma^{j}_{i})_{i=l}^{l_j-1},(\sigma^{j}_{i})_{i=l_j+1}^{n}\right)$}\; \lElse{$\sigma^{j,l}:=\left((\sigma^{r}_{i})_{i=1}^{l_r-1},(\sigma^{r}_{i})_{i=l_r+1}^{l},c,(\sigma^{r}_{i})_{i=l+1}^{n}\right)$}\;
	Guess $\sigma^{j,l}$\;
	$s:=\m{black}(\sigma^{j,l},y,x)$\;
	\If{$s>0$}
	{
	    $b:=l-1$\;
	    \lIf{$b<m$}{let $m:=b$}\;
	}
	\lElse{$a:=l$}\;
}
Return $m$\;
\caption{Function {\sc findNext}}\label{findNext}
\end{algorithm}

Let $j$ be an active index and $r$ its related index.
Let $c$ be the color of some component of $y$ that is already identified and fixed in the partial solution $x$.
With $l_{j}$ and $l_{r}$ we denote the position of color $c$ in $\sigma^{j}$ and $\sigma^{r}$ respectively.
The peg with color $c$ serves as a pivot peg for identifying a correct position $m$ in $\sigma^j$ that is not fixed, yet.
There are two possible modes for the binary search that depend on the fact if $m\le l_j$.
The mode is indicated by a boolean variable $\m{leftSearch}$ and determined by lines 4 to 8 of {\sc findNext}.
Clearly, $m\le l_j$ if $l_j=n$.
Otherwise, we guess
\[
\sigma^{j,0}:=\left(c,(\sigma^{j}_i)_{i=1}^{l_{j}-1},(\sigma^{j}_i)_{i=l_{j}+1}^{n}\right),
\]
By the information $\sigma^{j}_{i}=\sigma^{r}_{i+1}$ we obtain that $(\sigma^{j}_i)_{i=1}^{l_{j}-1}\equiv (\sigma^{r}_{i})_{i=2}^{l_{j}}$.
We further know that every open color has a wrong position in $\sigma^r$.
For that reason, $\m{black}(\sigma^{j,0},y,x)=0$ implies that $m\le l_j$.
The binary search for the exact value of $m$ is done in the interval $[a,b]$, where $m$ is initialized as $n$ and $[a,b]$ as
\[
[a,b]:=\begin{cases}
[1,l_j] & \text{if}\;\,\m{leftSearch}=\m{true}\\
[l_r,n] & \text{else}
\end{cases}
\]
(lines 9 to 11 of {\sc findNext}).
In order to determine if there is an open correct component on the left side of the current center $l$ of $[a,b]$ in $\sigma^j$ we can define a case dependent permutation:
\[
\sigma^{j,l}:=\begin{cases}
\left((\sigma^{j}_{i})_{i=1}^{l-1},c,(\sigma^{j}_{i})_{i=l}^{l_j-1},(\sigma^{j}_{i})_{i=l_j+1}^{n}\right) & \text{if}\;\,\m{leftSearch}=\m{true}\\
\left((\sigma^{r}_{i})_{i=1}^{l_r-1},(\sigma^{r}_{i})_{i=l_r+1}^{l},c,(\sigma^{r}_{i})_{i=l+1}^{n}\right) & \text{else}
\end{cases}
\]
In the first case, the first $l-1$ components of $\sigma^{j,l}$ coincide with those of $\sigma^j$.
The remaining components of $\sigma^{j,l}$ cannot coincide with the corresponding components of the secret code if they have not been fixed, yet.
This is because the $l$-th component of $\sigma^{j,l}$ has the already fixed value $c$, components $l+1$ to $l_j$ coincide with the corresponding components of $\sigma^r$ which satisfies $\m{black}(\sigma^r,y,x)=0$ and the remaining components have been checked to be wrong in this case.
Thus, there is a correct open component on the left side of $l$ in $\sigma^j$, if and only if $\m{black}(\sigma^{j,l},y,x)\ne 0$.
In the second case, the same holds for similar arguments.
Now, if there is a correct open component to the left of $l$, we update the binary search interval $[a,b]$ by $[a,l-1]$ and set $m:=\min(m,l-1)$.
Otherwise, we update $[a,b]$ by $[l,b]$.

\subsection{The Main Algorithm}
The main algorithm is outlined as Algorithm \ref{findAll}.
It starts with an empty partial solution and finds the components of the secret code $y$ one-by-one.
Herein, the vector $v$ does keep record about the number of open components in which the permutations $\sigma^1,\dots,\sigma^n$ equal $y$ and is, thus, initialized by $v_i:=\m{black}(\sigma^i,y)$, $i\in\{1,\dots,n-1\}$ and $v_n:=n-\sum_{i=1}^{n-1}v_i$.
As mentioned above, the main loop always requires an active index.
For that reason, if $v=\mathds{1}_n$ in the beginning, we fix one solution peg in $\sigma^1$ and update $x$ and $v$, correspondingly.
Every call of {\sc findNext} in the main loop augments $x$ by a correct solution value.
Since one call of $\m{findNext}$ requires at most $1+\lceil\log_2{n}\rceil$ guesses, Algorithm \ref{findAll} does not need more than $(n-3)\lceil\log_2{n}\rceil+\frac{5}{2}n-1$ queries (inclusive at most $\frac{n}{2}+1$ initial and $2$ final queries, respectively) to break the secret code.

\begin{algorithm}
Let $y$ be the secret code and set $x:=(0,0,\dots,0)$\;
Guess the permutations $\sigma^i$, $i\in\{1,\dots,n-1\}$ defined by (\ref{initialPermutations})\;
Initialize $v\in\{0,1,\dots,n\}^n$ by $v_i:=\mathrm{black}(\sigma^i,y)$, $i\in\{1,\dots,n-1\}$, $v_n:=n-\sum_{i=1}^{n-1}v_i$\;
\If{$v=\mathds{1}_n$}
{
    $j:=1$\;
	Find the position $m$ of the correct peg in $\sigma^1$ by at most $\frac{n}{2}+1$ further guesses\;
}
\Else
{
    Choose an active index $j\in\{1,\dots,n\}$ and call {\sc findFirst} to find the position of the correct peg in $\sigma^j$ by at most $2\lceil\log_2{n}\rceil$ further guesses\;
}
$x_m:=\sigma^j_m$\;
$v_j:=v_j-1$\;
\While{$|\{i\in\{1,\dots,n\}\,|\,x_i=0\}|>2$}
{
	Choose an active index $j\in\{1,\dots,n\}$\;
	$m:=\mbox{\sc findNext}(y,x,j)$\;
	$x_m:=\sigma^{j}_m$\;
	$v_{j}:=v_{j}-1$\;
}
Make at most two more guesses that are obtained from $x$ by assigning its two remaining zero-components to the two unidentified colors\;
\caption{Mastermind Algorithm for Permutations}\label{findAll}
\end{algorithm}

\section{More Colors than Components}
Now, we consider the case $k>n$ and forbidden color repetition.
Let $y=(y_1,\dots,y_n)$ be the code that must be found.
We use the same notations as above.

\subsection{Phase 1}
Consider the $k$ permutations $\overline{\sigma}^1,\dots,\overline{\sigma}^k$, where $\overline{\sigma}^1$ corresponds to the identity map on $\{1,\dots,k\}$ and for $j\in\{1,\dots,k-1\}$, we obtain $\overline{\sigma}^{j+1}$ from $\overline{\sigma}^{j}$ by a circular shift to the right.
We define $k$ codes $\sigma^1,\dots,\sigma^k$ by $\sigma^j=(\overline{\sigma}^j_i)_{i=1}^n$, $j\in\{1,\dots,k\}$,
i.e. we set
\begin{align*}
\sigma^1 &=(1,2,\dots,n),\\
\sigma^2 &=(2,\dots,n-1,n,n+1),\\
& \dots,\\
\sigma^{k-n+1} &=(k-n+1,k-n+2,\dots,k-1,k),\\
\sigma^{k-n+2} &=(k-n+2,k-n+3,\dots,k,1).\\
& \dots,\\
\sigma^{k-1} &=(k-1,k,1,\dots,n-3,n-2),\\
\sigma^{k} &=(k,1,\dots,n-1).
\end{align*}
Within those $k$ codes, every color appears exactly once at every position and, thus, we have
\[
\sum_{j=1}^{k}\m{black}(\sigma^j,y)=n,
\]
similar to (\ref{infosum}).
Since $k>n$, this implies that
\begin{lemma}
There is a $j\in \{1,\dots,k\}$ such that $\m{black}(\sigma^j,y)=0$.
\end{lemma}

\subsection{The Main Algorithm}
We are able to apply the routines described in Section \ref{section:KequalsN} by replacing the $n-1$ initial queries with the new $k-1$ queries $\sigma^{1},\dots,\sigma^{k-1}$ and initializing the vector $v$ correspondingly (lines 2 and 3 of Algorithm \ref{findAll}).
By Lemma 1, the case $v=\mathds{1}$ will not appear and we can always apply {\sc findFirst} to identify the first correct value.
For the required number of queries to break the secret code we have: The initial $k-1$ guesses, one call of {\sc findFirst} to detect the first correct position (at most $2\lceil\log_2{n}\rceil$ guesses), a call of {\sc findNext} for every other but the last two positions (at most $1+\lceil\log_2{n}\rceil$ guesses per position) and one or two final guesses.
This yields
\begin{theorem}
The modified Mastermind Algorithm for Permutations breaks the secret code in at most $(n-1)\lceil\log_2{n}\rceil+k+n-2$ queries.
\end{theorem}

\section{Conclusions and Further Work}

In this paper we presented a deterministic algorithm for the identification of a secret code in "Permutation Mastermind".
A challenge of Permutation Mastermind is that no color repetition is allowed for a query while most strategies for other Mastermind variants exploit the property of color repetition.
Further, concerning Permutation Mastermind, we consider for the first time the case that the number of colors is greater than the code length.
The provided Algorithms were implemented in Matlab and tested for $n,k\le 1000$.

In \cite{KT86} it is mentioned that a trivial lower bound on the worst case number of queries for Permutation Mastermind is $n$, but the authors conjecture that this number is actually $\Omega(n\log{n})$, a proof of which would close the gap to the upper bound.
The authors show that the search space reduces at most by a factor of $e$ after the first guess and point out that a similar assertion for the whole sequence of queries would yield the desired lower bound due to Stirling's formula for $n!$.
Experimental, we checked the worst case search space reduction after the first two queries to be less than $e^2$ for $n\le 9$.
Finding general theoretical formulas w.r.t. this matter remains a very ambiguous task.
However, further experiments may not only increase evidence about the lower bound conjecture but probably will even give ideas concerning expedient analysis steps.

[SODA 2013] Benjamin Doerr, Reto Sp\"{o}hel, Henning Thomas, and Carola Winzen.
Playing Mastermind with Many Colors
In: Proc. of ACM-SIAM Symposium on Discrete Algorithms (SODA 2013), pages 695-704, SIAM Society for Industrial and Applied Mathematics, 2013.
arXiv version.

\end{document}